\documentclass[12pt]{article}
\usepackage{pic02}
\usepackage{hyperref}
\usepackage{url}
\usepackage{graphicx}
\begin{document}
\title{\bf EXO: the Enriched Xenon Observatory for Double Beta Decay}
\author{K. Wamba\\
for the EXO Collaboration\\ 
{\em Stanford Linear Accelerator Center, Stanford, California, USA}}
\maketitle

\baselineskip=14.5pt
\begin{abstract}
EXO is a search for neutrinoless double beta decay in
\begin{math}^{136}\end{math}Xe. An active R\&D program for a 10 ton, enriched
\begin{math}^{136}\end{math}Xe liquid phase detector is now underway. 
Current research
projects are: decay product extraction, Xe  purity studies, energy
resolution studies, and Ba+ ion laser-tagging. Half lives of up to
5.0x10\begin{math}^{28}\end{math}yr will be ultimately probed,
corresponding to a sensitivity to Majorana neutrino masses
\begin{math}\geq 10\end{math}meV.
\end{abstract}

\baselineskip=17pt

\section{Introduction}

The EXO experiment is a search for
\begin{math}0\nu\beta\beta\end{math} in
\begin{math}^{136}\end{math}Xe, using several unique strategies to achieve
high sensitivity to small neutrino masses.  It consists of a
liquid Xe TPC in which we read out both the ionization and the
scintillation signals produced by double beta decays, thereby
obtaining the requisite energy resolution.  It also incorporates a
technique for laser tagging the daughter nucleus,
\begin{math}^{136}\end{math}Ba.
We refer the reader to an earlier publication by our collaboration
that describes in detail the theory of our detection scheme
\cite{dan}. Here, we give a status report on the R\&D program
currently underway that is leading to a 10 ton liquid
\begin{math}^{136}\end{math}Xe detector that incorporates this
novel approach.

\section{Xenon purification and  Ba ion extraction R\&D }

In our system optimum Xe purity is achieved by means of a SAES hot
Zr getter through which the Xe gas is passed, leaving behind any
substances that react with the hot Zr. Figure 1 shows a simplified
diagram of our setup.

\begin{figure}[htbp]
  \centerline{
    \includegraphics[width=7cm]{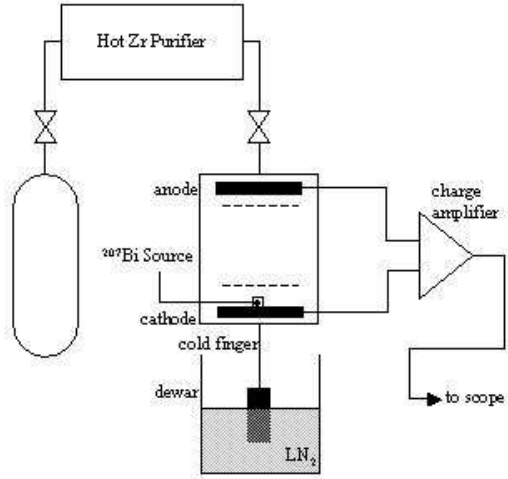}
  }
 \caption{\it
      Simplified diagram of the Xe purification and purity monitoring setup.
    \label{system} }
\end{figure}

Using an adaptation of a technique first demonstrated by Carugno,
et al. \cite{car}, we measure the resulting purity by condensing
the Xe into a test cell and drifting electrons through it.  Our
test cell holds about 100cc of liquid Xe and contains an anode,
field shaping electrodes and a cathode with a
\begin{math}^{207}\end{math}Bi source attached to it. 
Using an adaptation of a technique first demonstrated by Carugno,
et al. \cite{car}, we measure the resulting purity by condensing
the Xe into a test cell and drifting electrons through it.  Our
test cell holds about 100cc of liquid Xe and contains an anode,
field shaping electrodes and a cathode with a
\begin{math}^{207}\end{math}Bi source attached to it.  The ionization
charge liberated by the radioactivity from the source is drifted
across the 6cm length of the cell. The electron lifetime is
determined by comparing the charge signal amplitude when it leaves
the cathode to when it arrives at the anode. Figure 2 shows a
diagram of our purity cell and a sample oscilloscope trace of the
shaped output from our charge sensitive amplifier. In the scope
trace, the positive pulse corresponds to the signal induced by
electrons leaving the cathode; the negative pulse, electrons
arriving at the anode. Note
that only a very slight decrease in pulse amplitude is apparent.
\begin{figure}[htbp]
  \centerline{\hbox{ \hspace{0.2cm}
    \includegraphics[height=6cm]{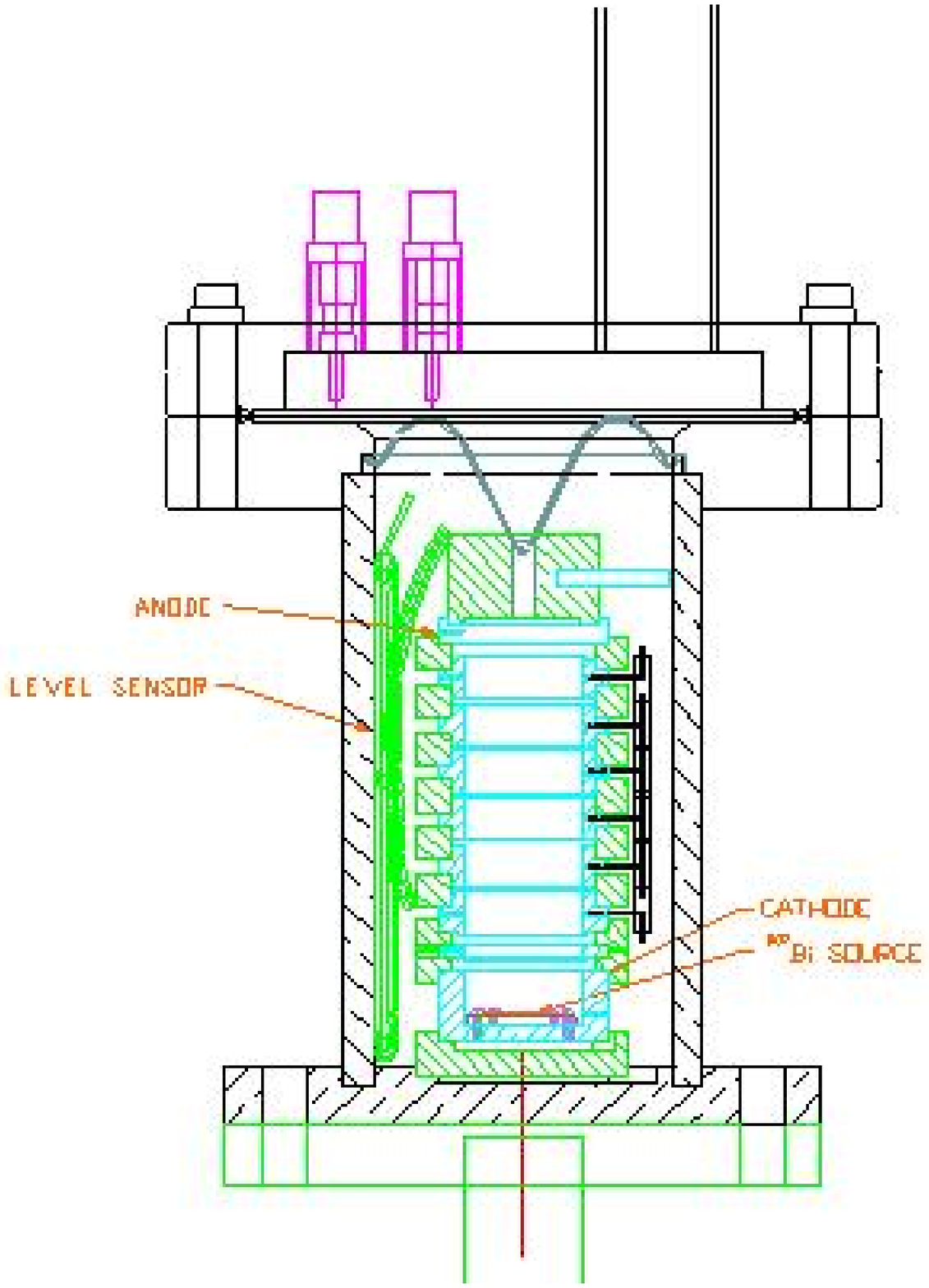}
    \hspace{0.1cm}
    \includegraphics[height=5cm]{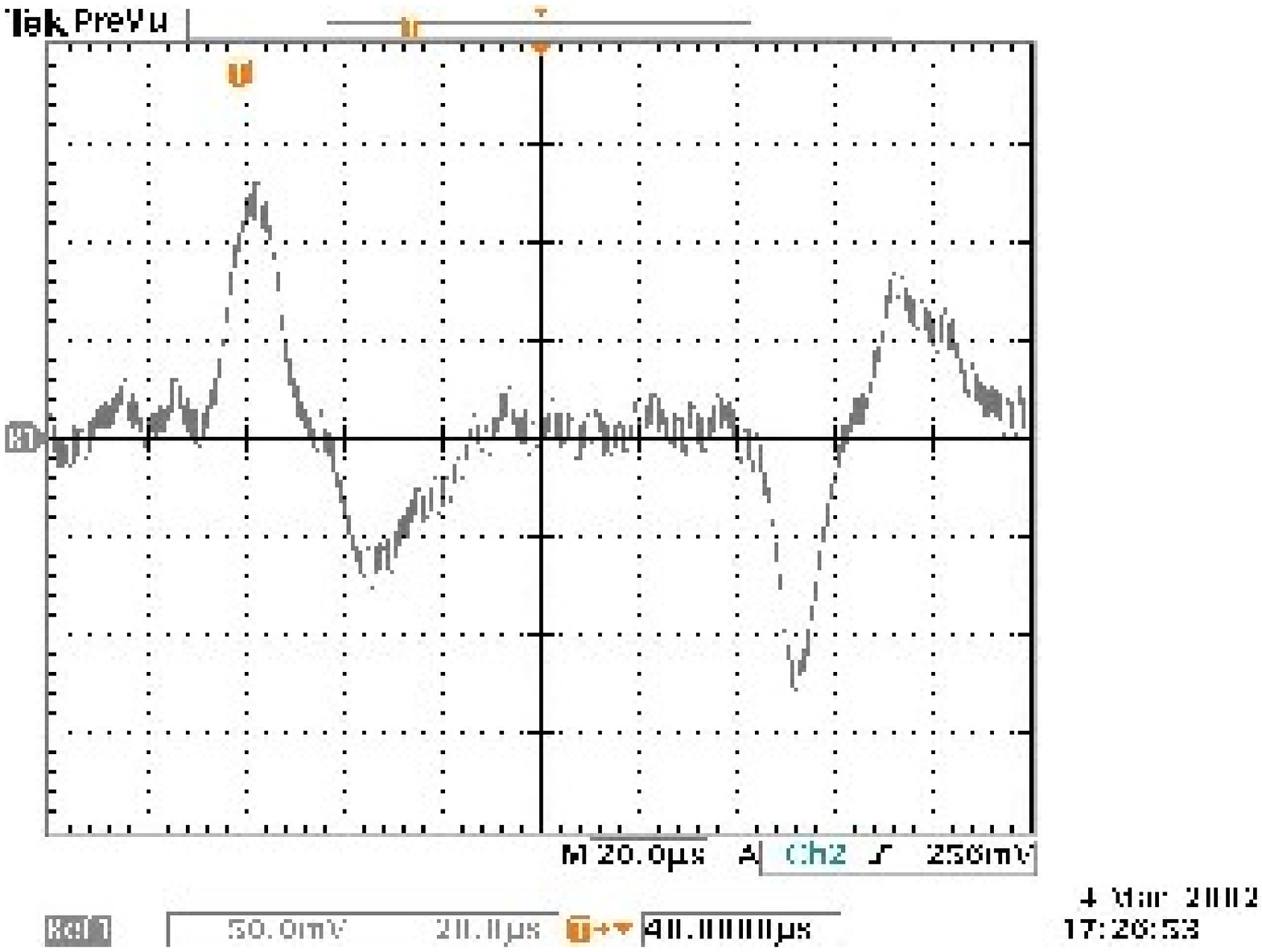}
    }
  }
 \caption{\it
  Left:  purity monitor cell for measuring electron lifetime in liquid Xe.  
  Right: Oscilloscope trace of one ionization event.
    \label{xpm} }
\end{figure}


We are also investigating the possibility of extracting the Ba ion
from the liquid and releasing it
 into an analysis chamber for identification.  We have developed a
 special cell in which a small amount of liquid Xe  is condensed.  
A \begin{math}^{230}\end{math}U source emits \begin{math}^{222}\end{math}Ra ions
 into the liquid.  We then attempt to extract these ions
  from liquid solution by means of an electrostatic probe, thereby
   proving the overall principle of individual ion extraction.
Our initial tests are currently being performed with a bare
tungsten tip that is held at a negative high voltage in order to
collect ions from solution. We have just begun initial testing of this system.

\section{Conclusion}

The remainder of our R\&D program includes a setup for the optimization and study of the energy and position resolution achievable in liquid Xe and an experiment to determine the viability of trapping and spectroscopic study of the Ba\begin{math}^+\end{math} and the Ba\begin{math}^{++}\end{math} ion.  We expect to publish initial results from these activities in the near future.
Once all of the various techniques are brought together in an
ultimate 10-ton experiment, we expect to be sensitive to half
lives as long as 5.0x10\begin{math}^{28}\end{math}yr, or an
effective neutrino mass in the range 10-50meV.  This is about a
factor of 10 below the world's current best value, as reported in
\cite{gue}.

\end{document}